\documentclass[nojss]{jss}

\usepackage{thumbpdf,lmodern}

\newcommand{\class}[1]{`\code{#1}'}
\newcommand{\fct}[1]{\code{#1()}}

\author{Unai P{\'e}rez-Goya\\Public University of Navarre\\InaMat Institute
   \And Manuel Montesino-SanMartin\\Public University of Navarre\\InaMat Institute
   \AND Ana F. Militino\\Public University of Navarre\\InaMat Institute
   \And M. Dolores Ugarte\\Public University of Navarre\\InaMat Institute}
\Plainauthor{Unai P{\'e}rez-Goya, Manuel Montesino-SanMartin, Ana F. Militino, M. Dolores Ugarte}

\title{\pkg{RGISTools}: Downloading, Customizing, and Processing Time-Series of
Remote Sensing Data in \proglang{R}}
\Plaintitle{RGISTools: Downloading, Customizing, and Processing Time 
Series of Remote Sensing Data in R}
\Shorttitle{\pkg{RGISTools}: Time-Series of Remote Sensing Data in \proglang{R}}

\Abstract{
There is a large number of data archives and web services offering free
access to multispectral satellite imagery. Images from multiple sources are
increasingly combined to improve the spatio-temporal coverage of measurements
while achieving more accurate results. Archives and web services differ in
their protocols, formats, and data standards, which are barriers to combine
datasets. Here, we present \pkg{RGISTools}, an \proglang{R} package to create
time-series of multispectral satellite images from multiple platforms in a
harmonized and standardized way. We first provide an overview of the package
functionalities, namely downloading, customizing, and processing multispectral
satellite imagery for a region and time period of interest as well as a recent
statistical method for gap-filling and smoothing series of images, called
interpolation of the mean anomalies. We further show the capabilities of the 
package through a case study that combines Landsat-8 and Sentinel-2 satellite
optical imagery to estimate the level of a water reservoir in Northern Spain.
We expect \pkg{RGISTools} to foster research on data fusion and spatio-temporal
modelling using satellite images from multiple programs. 
}

\Plainkeywords{Landsat, MODIS, Sentinel, satellite images, spatio-temporal data,
IMA}

\Address{
  Unai P{\'e}rez-Goya\\
  Department of Statistics, Computer Science, and Mathematics\\
  Public University of Navarre\\
  31006 Pamplona, Spain\\
  E-mail: \email{unai.perez@unavarra.es}\\
  URL: \url{https://www.unavarra.es/pdi?uid=811058}\\
  
  Manuel Montesino-SanMartin\\
  Department of Statistics, Computer Science, and Mathematics\\
  Public University of Navarre\\
  31006 Pamplona, Spain\\
  E-mail: \email{manuel.montesino@unavarra.es}\\
  
  Ana F. Militino\\
  Department of Statistics, Computer Science, and Mathematics\\
  Public University of Navarre\\
  31006 Pamplona, Spain\\
  E-mail: \email{militino@unavarra.es}\\
  URL: \url{https://www.unavarra.es/personal/amilitino}\\
  
  Maria Dolores Ugarte\\
  Department of Statistics, Computer Science, and Mathematics\\
  Public University of Navarre\\
  31006 Pamplona, Spain\\
  E-mail: \email{lola@unavarra.es}\\
  URL: \url{https://www.unavarra.es/personal/lugarte}\\
}

\begin{document}



\section[Introduction]{Introduction} \label{sec:intro}

Satellite images represent a valuable data source in large-scale long-term
research studies. Landsat, MODIS, and Copernicus are major programs for the 
acquisition of images of the Earth's surface supported by the U.S. Geological
Survey (USGS), NASA, and the European Space Agency (ESA) respectively. Images
are freely accessible in large  data archives, which can be retrieved via web
services such as EarthData, NASA Inventory or SciHub. Data archives offer long
series of records, dating back to 1972 for Landsat, 1999 for MODIS and 2013 for
Sentinel. Satellite imagery has proven useful for studies in many disciplines,
such as poverty assessments \citep{jean:2016}, glacier dynamics 
\citep{paul:2016}, soil classification \citep{gomez:2019}, distribution of
animal species \citep{swinbourne:2018}, and crop monitoring \citep{azzari:2017}.  

Missions have strengths and weaknesses regarding the spatial and temporal 
resolution of their imagery. The satellite constellation of MODIS acquires
images on a daily basis at a moderate spatial resolution (250m). Landsat and
Sentinel multispectral constellations capture high-resolution images (15-60m
and 10-60m respectively) where locations are revisited roughly on a weekly basis
(8 and 5 days). Studies claim the need for a higher spatio-temporal resolution
than those obtained from single programs \citep{griffiths:2019}. Data fusion
has been proposed to counteract inadequate resolutions by blending information
at different levels, pixel-level (e.g., MODIS and Sentinel), feature-level
(e.g., class of land-cover) or the decision-level \citep{belgiu:2019}. This is
partly possible thanks to improvements in availability and accessibility of
satellite images over the last decade. Some challenges still remain. Web
services and programs work with particular query protocols, file formats, and
data standards. Becoming familiar with the details of every archive can be
tedious and time consuming. A harmonized single access point and processing
software would benefit the research community removing complexity and fostering
data fusion.

\proglang{R} \citep{renv:2019} is an open source software increasingly used for
the analysis of satellite images, as it enables the application of 
state-of-the-art statistical methods. There are many reliable packages to
manipulate spatial or spatio-temporal data, such as \pkg{raster}
\citep{raster:2019} and \pkg{sf} \citep{sf:2018}, or to perform spatio-temporal
statistical analyses, such as \pkg{gstat} \citep{gstat:2004}. Packages working
with satellite images already exist in \proglang{R}. Few packages deal with 
imagery from several programs, but they are focused on specific tasks of the
overall workflow with satellite images. \pkg{SkyWatchr} \citep{skywathcr:2017}
finds and downloads Landsat, MODIS, Sentinel, and private company's imagery but
does not support data processing or customization. \pkg{ASIP} \citep{asip:2018}
is able to carry out a restricted set of processing steps for Landsat and 
Sentinel imagery, such as atmospheric corrections and spectral index 
computations, leaving uncovered cloud masking or smoothing. Other packages have
greater functionalities but they are specialized in particular programs or data
products. For instance, \pkg{MODIStsp} \citep{modistsp:2016} downloads, mosaics,
re-projects, and computes spectral indices from MODIS images exclusively.
\pkg{MODIS} \citep{modis:2019} and \pkg{MODISTools} \citep{modistools:2014} also
work with MODIS imagery but with more restricted functionalities. \pkg{MODISnow}
\citep{modissnow:2016} and \pkg{modiscloud} \citep{modiscloud:2013} only access
snowcover products and cloud masks, respectively. Regarding Sentinel-2, the
\pkg{sen2r} package \citep{sen2r:2019} is capable of finding, downloading,
and processing data products just from this satellite mission. The \proglang{R}
packages \pkg{landsat} \citep{landsat:2011}, \pkg{satellite} 
\citep{satellite:2015}, and \pkg{landsat8} \citep{landsat8:2017} mainly perform 
radiometric and topographic corrections of Landsat (or Landsat-8), but they
are not able to do the download. Consequently, there is a need for a 
comprehensive package that harmonizes the work with different satellite
programs.

\pkg{RGISTools} \citep{rgistools:2019} is conceived in response to those needs.
The package is a toolbox to work with time-series of satellite images from 
Landsat, MODIS, and Sentinel repositories in a standardized way. The functions
of \pkg{RGISTools} allow to build a semiautomatic line of work for downloading,
customizing, and processing imagery. The download process includes the search
and preview of images for a region and period of interest. The customization 
covers image mosaicking, cropping, and extracting the required bands. Processing
functions comprise cloud removal, definition of new variables, gap filling, and
image smoothing. \pkg{RGISTools} is available from the Comprehensive \proglang{R}
Archive Network in \url{https://cran.r-project.org/web/packages/RGISTools/index.html}
and the Git hub repository in \url{https://github.com/spatialstatisticsupna/RGISTools}.

The structure of this paper is as follows: Section~\ref{sec:programs} introduces
basic information to handle satellite images. Section~\ref{sec:overview} gives
an overview of the work sequence with the package. This section provides brief
descriptions of the aim and inputs of each function. Explanations are coupled
with a MODIS example on using the interpolation of the mean anomalies (IMA) 
procedure for gap-filling and smoothing images that is available in the package.
In Section~\ref{sec:example}, we present an example that combines Landsat-8
and Sentinel-2 to monitor the water levels of a reservoir in Northern Spain.



\section{Satellite programs} \label{sec:programs}

The package focuses on optical imagery, which is the form of satellite 
information most commonly used in research. Operational satellite missions
concerning with optical measurements are Landsat-7, Landsat-8, MODIS, and
Sentinel-2.

\subsection{Data types and structure} \label{subsec:data}

\subsubsection{Wavelengths and band names} \label{subsubsec:bands}

The type and structure of satellite data varies with the mission. Each 
mission involves one or several satellites that carry purpose-specific 
instruments (Table~\ref{tab:satsummary}). On board instruments measure the
solar radiance in specific bands of the electromagnetic spectrum. For instance,
the Terra and Aqua satellites from MODIS carry on-board the moderate resolution
imaging spectroradiometer (MODIS). It captures 36 bands in the visible and
infrared parts of the spectrum \citep{MODISweb}. MODIS collects information on
a greater number of bands and with narrower spectral windows than Landsat-7 (8 
bands), Landsat-8 (11 bands) \citep{Landsatweb}, and Sentinel-2 (12 bands)
\citep{Sentinelweb} satellites. Bands are identified by numbers, which are
given in sequential order. Similar wavelengths might be labelled with different
numbers depending on the mission. For instance, the red band 
($0.673-0.695 \mu m$) is the band 3 in Landsat-7's imagery, the band 4 in
Landsat-8's and Sentinel-2, and bands 1, 13, and 14 in MODIS. Computing remote
sensing indices can be problematic due to inconsistencies in the band names.

\subsubsection{Tiling systems} \label{subsubsec:tiles}

Satellite records are partitioned into scenes that cover portions of the earth's
surface, called tiles. Tiling systems are conceived to facilitate data processing
and sharing. Each mission has its own tiling system, varying in tile's size,
orientation, and naming conventions. For example, MODIS tiles are considerably
larger ($1200 \times 1200 \, km^{2}$ ) than the ones used for Landsat-7 
($170 \times 183 \, km^{2}$), Landsat-8 ($185 \times 180 \, km^{2}$) or 
Sentinel-2 ($100 \times 100 \, km^{2}$). Satellite programs provide keyhole
markup language files (KML) with the boundaries of the tiles at their respective
official websites \citep{Landsatkml, MODISkml, Sentinelkml}. Depending on the
mission, one or several tiles can cover the region of interest. In the latter
situation, images should be properly merged and cropped.

\subsubsection{Data products and processing levels} \label{subsubsec:products}

Sensor features, radiometric, and geometric effects distort satellite images.
Corrections are required to convert sensor data into surface reflectances.
Programs offer several products depending on the level of processing being
applied. Generally, level-2 products are processed to provide the surface
reflectances and they are suitable for most applications. MODIS additionally
distinguishes different products depending on the scientific field to which
the information is targeted (atmospheric, cryogenic, and land products)
\citep{mod:dataprod}. Based on the purpose of the satellite imagery, the
researcher must select the appropriate product and processing level. During the
correction process, images are also geo-referenced. MODIS defines the coordinates
of the pixels using the global sinusoidal projection \citep{MODISweb}, while
Landsat and Sentinel use the universal trade mercator (UTM) system under the
world geodetic system 1984 (WGS84) \citep{Landsatweb, Sentinelweb}. Any fusion
between MODIS and Landsat/Sentinel datasets would require to re-project one of
two collection of images.

\subsection{Sharing protocols and data formats} \label{subsubsec:sharing}

\subsubsection{Web services} \label{subsubsec:webservices}

Web services represent an interactive mean to access the archives of one or
several programs. They offer one or two ways to access the imagery: through a
graphic user interface (GUI) or an application programming interface (API). APIs
are specially convenient to search and download time-series of satellite images
programatically. Major existing web services with APIs are EarthData 
\citep{Earthdata}, NASA Inventory \citep{NASAinventory}, and SciHub 
\citep{Scihub}. Users can select among several query options and should
interpret the response in extensible markup language (XML) or javascript 
object notation (JSON).

\subsubsection{Formats} \label{subsubsec:formats}

Pixel values are re-scaled and images are compressed to preserve the information
efficiently and accurately. Satellite programs use different formats and
compression methods (see Table~\ref{tab:satsummary}). Landsat images are encoded
as GTiff and stored as tape archive files (".tar") and GNU compression 
standards (".gz") \citep{Landsatweb}. MODIS images are shared in hierarchical
data format (".hdf") \citep{mod:dataprod}. Sentinel images are available as
raster images using JPEG2000 format (".jp2") and encapsulated as ".tar.gz" files
\citep{Sentinelkml}. Images must be extracted and once imported, pixel values
representing surface reflectance are usually scaled between 0 and 10000.
However, actual ranges are generally larger as a result of the correction
algorithms. In MOD09GA, surface reflectance goes from -100 to 16000. Pixel
values should be truncated and re-scaled for some applications.

\begin{table}[t!]
\centering
\begin{tabular}{l|c|c|c|c|c|c}
\hline
Program             & \multicolumn{2}{c|}{Landsat} & \multicolumn{2}{c|}{MODIS} & \multicolumn{2}{c}{Sentinel} \\ \hline
Mission             & Landsat-7 & Landsat-8        & \multicolumn{2}{c|}{-}     & \multicolumn{2}{c}{Sentinel-2} \\ \hline 
Satellite           & Landsat-7 & Landsat-8        & Terra   & Aqua             & A          & B \\
Sensor               & ET+       & TIRS/OLI         & MODIS   & MODIS            & MSI        & MSI \\
No. Bands           & 8         & 8                & 36      & 36               & 12         & 12 \\
Time Revisit (days) & 16        & 16               & 1       & 1                & 10         & 10 \\
Resolution (m)      & 30-60     & 15-30            & 250     & 250              & 10-60      & 10-60 \\
Format              & GTiff   & GTiff          & HDF-EOS & HDF-EOS          & JP2        & JP2 \\ \hline
\end{tabular}
\caption{\label{tab:satsummary} Major satellite missions devoted to multi-spectral
images and details about their datasets.}
\end{table}

The aim of \pkg{RGISTools} is to centralize the information, standardize, and
automate satellite imagery retrival, customization, and processing. The following
sections describe how to use the package to obtain a complete and ready-to-use
time-series of remote sensing data.

\section[RGISTools overview]{\pkg{RGISTools} overview} \label{sec:overview}

The \pkg{RGISTools} package works with multiple sources of information and, for
this reason, the functions are grouped into 5 categories depending on the 
mission they focus on. Functions begin with one of the following prefixes:

\begin{itemize}
  \item \verb|ls|, \verb|mod|, and \verb|sen| involve Landsat, MODIS and 
  Sentinel imagery respectively. More specifically, \verb|ls7| and \verb|ls8|
  are restricted to Landsat-7 and Landsat-8 missions.
  \item \verb|gen| can be applied to images from any mission.
  \item \verb|var| compute widespread remote sensing indices.
\end{itemize}

The package implements a variety of procedures related to downloading,
customizing, and processing satellite images. A suffix in the function's name
indicates its purpose. The main functionalities of \pkg{RGISTools} are
introduced in the following sections along with an example analysing the 
spatio-temporal evolution of the Normalized Difference Vegetation Index (NDVI)
\citep{ndvi:1972}.

\pkg{RGISTools} downloads and works with satellite imagery locally on
your computer. Then, as a memory-saving strategy, most functions deal with
images externally to \proglang{R}. The workflow is designed to delay the data
loading in the \proglang{R} environment until the end of the customization. At
this point, the relevant data have been transformed to meet the particular 
needs of the analysis. As a result, rather than \proglang{R} objects, 
downloading and customizing functions take a file path as an input (\code{src}
argument) and generate GTiffs and folders in a given directory (\code{AppRoot}
argument) as an output. Functions print a message when completing their task
to help remembering the output location. A clear hierarchical structure of
folders and an appropriate file management are key to work successfully with
\pkg{RGISTools}.

The NDVI example requires in total $0.92$ Giga Bytes (GB) of memory space. It
takes nearly 5 minutes to run from top to bottom in an intel(R) Core(TM) 
i7-6700 CPU @3.40 GHz and an internet connection speed of $310$ Mbps. In case of 
insufficient memory space, we provide links throughout the next sections to
download the resulting files. After data processing, the file size decreases
from a maximum of $198$ MB to a minimum of $3$ MB.

\subsection{Retrieving satellite imagery} \label{subsec:retrieve}

Retrieving satellite imagery involves three steps; searching, previewing, and
downloading scenes for a specific time-period and region of interest (ROI).
Some of these steps require valid credentials from EarthData \citep{earthdata:reg}
and SciHub \citep{scihub:reg} web services, which can be acquired after 
registration in their respective websites.

\subsubsection{Searching} \label{subsubsec:search}

The first step in retrieving satellite images is to search the scenes available
for a particular ROI and time window. Search results provide valuable
information on the number of available images, the dates they were captured, or
the tiles they belong to. The \fct{lsSearch}, \fct{modSearch}, and 
\fct{senSearch} functions require as inputs the name of the data product, the
time interval, and the ROI.

A data product is a collection of images with certain bands and processing level.
Products are identified by short-names, which can be found in Landsat, MODIS,
and Sentinel websites and product guides \citep{mod:dataprod, sen:dataprod}.
The spatio-temporal domain under analysis is specified through a time interval
(\code{dates}) and a location (\code{region}). The time span is defined by a
vector of \class{Date} class objects and the ROI can be any spatial object in
\proglang{R}  (\class{Spatial*}, \class{sf}, or \class{raster}).

In the following, we search multispectral images of the surface reflectance
(level-2) of optical bands captured by the Terra satellite (``MOD09GA'' product)
between the $2^{nd}$ and $9^{th}$ of August 2018. The ROI is the Navarre
province located in Northern Spain. The border of this region is represented
in \code{ex.navarre} as a \class{SimpleFeature} with a \class{MULTIPOLYGON}
geometry:

\begin{Schunk}
\begin{Sinput}
R> library("RGISTools")
R> wdir <- tempdir()
\end{Sinput}
\end{Schunk}
\begin{Schunk}
\begin{Sinput}
R> data("ex.navarre")
R> sres <- modSearch(product = "MOD09GA",
+                    dates = as.Date("2018-08-02") + seq(0 , 7, 1),
+                    region = ex.navarre)
\end{Sinput}
\end{Schunk}

\subsubsection{Previewing} \label{subsubsec:preview}

The second step of retrieving satellite imagery is previewing the search results.
Previewing might be useful to inspect the spatial coverage and cloudiness of the
imagery. Thus, some images can be discarded at an early stage, saving time during
the download and image processing. The functions \fct{lsPreview}, \fct{modPreview},
and \fct{senPreview} display a color picture of an image on a map in the viewer
of RStudio. The images being displayed are the ones captured on a given date
(\code{dates}). The map allows to zoom-in and -out to preview in an appropriate
level of detail.

\begin{figure}[t!]
   \centering
   \includegraphics[height=75mm,keepaspectratio]{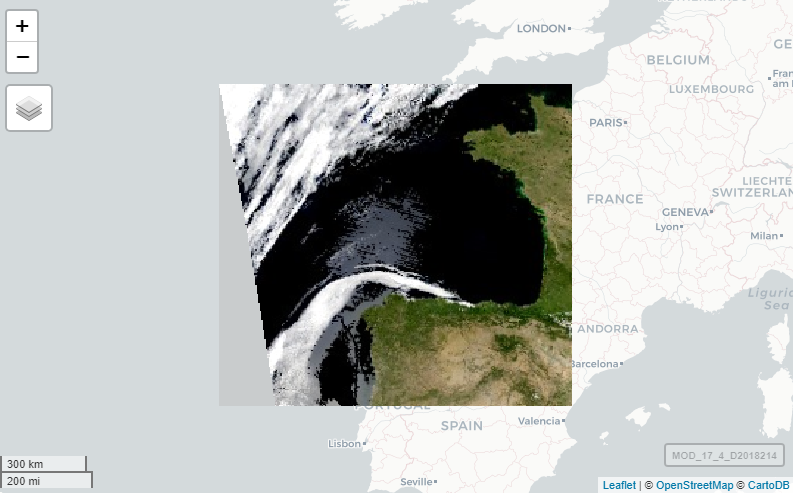}
   \caption{\label{fig:preview} A preview of the $1^{st}$ image of the ``MOD09GA''
   time-series. The image corresponds to the ``h:17v:4'' tile from MODIS, which
   covers the region of Navarre (\texttt{ex.navarre}) in Northern Spain. The
   image was captured on August $2^{nd}$, 2018 by the Terra satellite.}
 \end{figure}

The following code displays the preview of the $1^{st}$ element in
\code{searchres_preview} (Figure~\ref{fig:preview}):

\begin{Schunk}
\begin{Sinput}
R> modPreview(searchres = sres, dates = as.Date("2018-08-02"))
\end{Sinput}
\end{Schunk}

\subsubsection{Downloading} \label{subsubsec:download}

The functions \fct{lsDownload}, \fct{modDownload}, or \fct{senDownload} download
and uncompress satellite images from a search list (\code{searchres}). The user
can specify the folder where the imagery will be placed using the \code{AppRoot}
argument or images will be saved in the current working directory otherwise. 

The function downloads and saves the satellite images in their original format
in a folder automatically created under \code{AppRoot}. If the proper flag is
active (e.g., \code{extract.tif = TRUE} in MODIS), the function decompresses and
transforms the imagery to GTiff. The uncompressed images are saved in
another folder also generated automatically in \code{AppRoot}. If only few bands
of the spectrum are needed, the argument \code{bFilter} allows to specify which
bands should be transformed.

Below, we download and uncompress the previously found time-series of images
(\code{sres}). As mentioned earlier, the imagery will be used to compute the
NDVI index (see Section~\ref{subsec:customize}), so the red (``B01'') and
near-infrared (``B02'') bands must be extracted. We also require the quality
band (``state'') to be able to remove the pixels covered by clouds. 

To run the next code, replace the \verb|<USERNAME>| and \verb|<PASSWORD>| with
the credentials acquired at \cite{earthdata:reg}. Images are saved in the
\code{wdir.mod.download} directory (i.e., \verb|./Modis/MOD09GA|) inside a
temporary directory:

\begin{Schunk}
\begin{Sinput}
R> wdir.mod <- file.path(wdir, "Modis")
\end{Sinput}
\end{Schunk}
\begin{Schunk}
\begin{Sinput}
R> wdir.mod <- file.path(wdir, "Modis")
R> wdir.mod.download <- file.path(wdir.mod, "MOD09GA")
R> modDownload(searchres = sres,
+              AppRoot = wdir.mod.download,
+              extract.tif = TRUE,
+              bFilter = c("B01", "B02", "state"),
+              username = "<USERNAME>",
+              password = "<PASSWORD>",
+              overwrite = TRUE)
\end{Sinput}
\end{Schunk}

The preview might not be necessary when further filtering is not required or
there is no interest in exploring the tiles covering the ROI. In these
situations, the functions \fct{lsDownSearch}, \fct{modDownSearch}, and 
\fct{senDownSearch} can search, download, and uncompress the time-series of
images at once. An example follows:

\begin{Schunk}
\begin{Sinput}
R> modDownSearch(product = "MOD09GA",
+                dates = as.Date("2018-08-02") + seq(0, 7 , 1),
+                region = ex.navarre,
+                AppRoot = wdir,
+                extract.tif = TRUE,
+                bFilter = c("B01", "B02", "state"),
+                username = "<USERNAME>",
+                password = "<PASSWORD>")
\end{Sinput}
\end{Schunk}

The code above takes few minutes to run and requires $0.913$ GB of space in
the disk. The user can download the results as GTiff files ($0.198$ GB) from 
the reference \cite{vermonte:2019:dw}. Please, unzip the file and save it in
the \verb|./Modis| folder to continue with the example.

\subsection{Customizing satellite imagery} \label{subsec:customize}

Here, customizing satellite images refers to mosaicking, cropping, and computing
remote sensing indices. 

\subsubsection{Mosaicking and cropping} \label{subsubsec:mosaic}

Mosaicking means joining satellite images captured on the same date and from
different tiles to obtain a single scene covering the ROI. Cropping is the
removal of pixels outside the spatial bounding box that encapsulates the ROI.
Both tasks are meant to rearrange the dataset and preserve the relevant 
information only. Mosaicking and cropping functions are named after the
corresponding satellite mission and the keyword \verb|Mosaic| (i.e., 
\fct{lsMosaic}, \fct{modMosaic}, and \fct{senMosaic}). These functions require
the path to the folder that contains the uncompressed image files (\code{src}).
When provided, the function crops the image around the bounding box of the
spatial object (\class{Spatial*}, \class{sf}, or \class{raster}) that is passed
through the argument \code{region}.

Mosaic functions use by default the Geospatial Data Abstraction Library 
\citep{gdal:2019} through the the \pkg{sf} package interface \citep{sf:2018}.
If \code{gutils} is set to \code{FALSE}, the function borrows the mosaic
functionalities from the \pkg{raster} package \citep{raster:2019}. However, 
GDAL is more computationally efficiently than \pkg{raster}. The results
are saved in a new folder in the \code{AppRoot} directory named as the
\code{out.name} argument.

Mosaicking and cropping the imagery from previous examples is shown below.
Cropped images are saved into a folder called \verb|Navarre| under the
\code{wdir.mod} directory (i.e., \verb|./Modis|):

\begin{Schunk}
\begin{Sinput}
R> wdir.mod.tif <- file.path(wdir.mod,"MOD09GA","tif")
R> modMosaic(src = wdir.mod.tif,
+            region = ex.navarre,
+            out.name = "Navarre",
+            gutils = TRUE,
+            AppRoot = wdir.mod)
\end{Sinput}
\end{Schunk}

The MODIS tile covering Navarre is unique (``h17:v4''), so in our example,
\fct{modMosaic} just crops the images around the bounding box of 
\code{ex.navarre}.

Mosaicking and cropping takes few seconds to run with \code{gutils = TRUE}.
The size of the overall outcoming images is $3.72$ MB. To ensure that the rest
of the analysis is reproducible, the results are available at the reference
\cite{Vermonte:2019:mos}. No more files are provided through links hereinafter
for the MODIS example, as we consider that the size of the data set is
manageable, and the computational times for the rest of the example are sensible.

\subsubsection{Computing remote sensing indices} \label{subsubsec:index}

A common use of multispectral images is the computation of remote sensing indices.
These are mathematical expressions combining the reflectance of several bands
of the spectrum to highlight the phenomenon under analysis. The package includes
pre-built functions that define widespread remote sensing indices (i.e.,
\fct{varNDVI}, \fct{varEVI}, \fct{varNBR}, etc.). The Normalized Difference
Vegetation Index (NDVI) \citep{ndvi:1972} is a commonly used index to monitor
green vegetation. It uses the red and near-infrared wavelengths \citep{ndvi:2015}
due to the high levels of absorption and reflection in these wavelengths by
plants.

The functions \fct{lsFolderToVar}, \fct{modFolderToVar}, and \fct{senFolderToVar}
apply the \verb|var| functions over a time-series of multispectral satellite
images. The family of \verb|FolderToVar| functions requires as inputs the path
to the folder that stores the mosaicked images (\code{src} argument) and the
function that computes the remote sensing index (\code{fun} argument). The
outputs are saved in a folder named after the remote sensing index, in the
\code{AppRoot} directory.

For instance, the following code calculates a daily series of NDVIs from the
images mosaicked in the previous section. The resulting images are saved in
\code{wdir.mod} (i.e., \verb|./Modis/NDVI|):

\begin{Schunk}
\begin{Sinput}
R> wdir.mod.mosaic <- file.path(wdir.mod, "Navarre")
R> modFolderToVar(src = wdir.mod.mosaic,
+                 fun = varNDVI,
+                 AppRoot = wdir.mod)
\end{Sinput}
\end{Schunk}

\begin{figure}[t!]
 \centering
\includegraphics{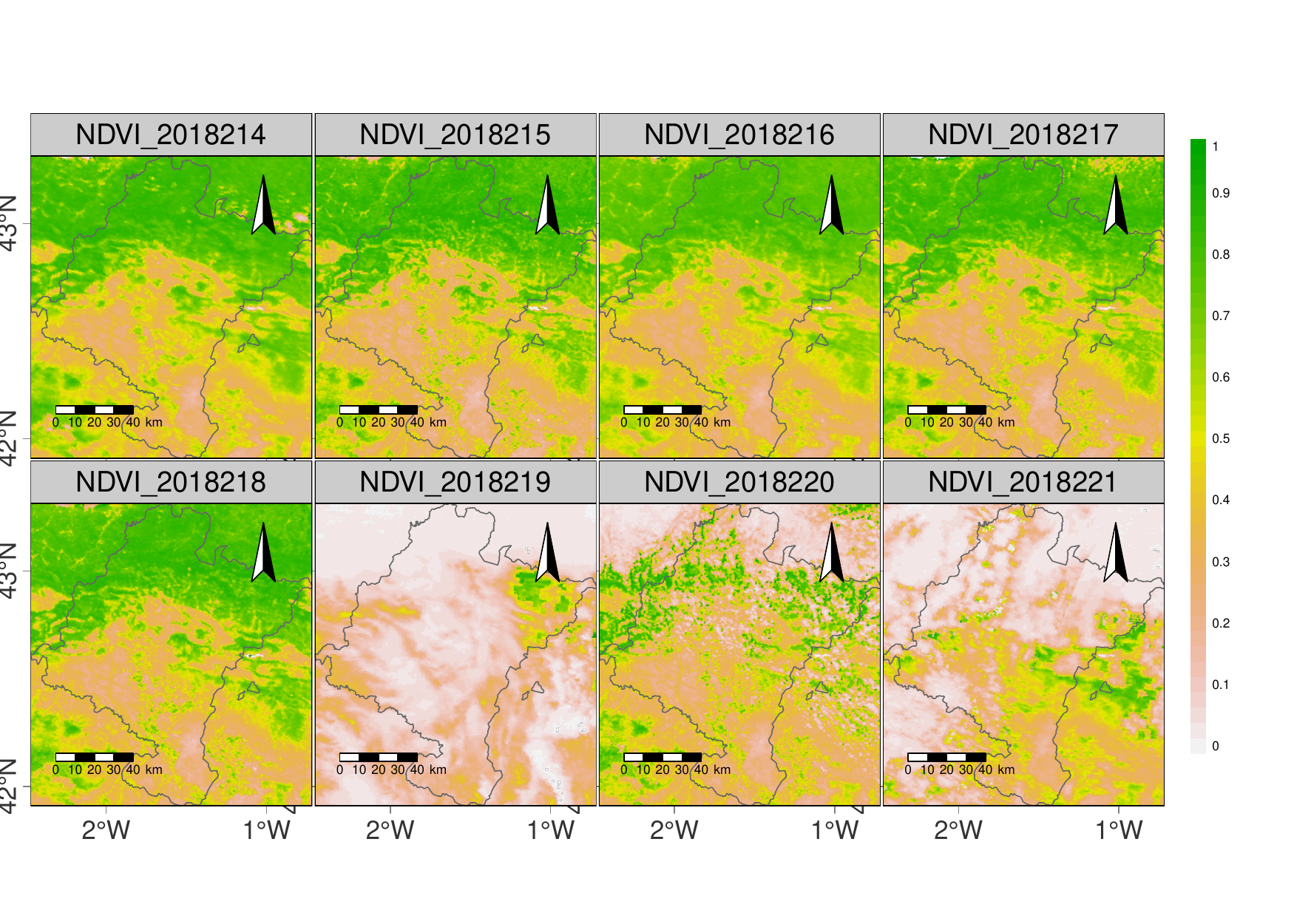}
 \vspace*{-10mm}
 \caption{\label{fig:ndvifig} Time-series of images showing the NDVI in Navarre
 between the $2^{nd}$ and $9^{th}$ of August, 2019. The dates in the panels are
 in \texttt{YYYYJJJ} format, where \texttt{Y} is a year digit and \texttt{J} is a
 Julian day digit. The line represents the border of the region of Navarre.}
\end{figure}

The generated data can be loaded in \proglang{R} using the \fct{stack} function
from the \pkg{raster} package \citep{raster:2019} (Figure~\ref{fig:ndvifig}).
Due to errors in some pixel values, results of the NDVI may yield results 
outside the usual -1 and 1 range \citep{ndvi:1972}. These artifacts can be
removed with the function \fct{clamp} from \pkg{raster} \citep{raster:2019} as 
follows:

\begin{Schunk}
\begin{Sinput}
R> wdir.mod.ndvis <- file.path(wdir.mod, "NDVI")
R> files.mod.ndvi <- list.files(wdir.mod.ndvis, full.names = TRUE)
R> imgs.mod.raw  <- raster::stack(files.mod.ndvi)
R> imgs.mod.ndvi <- raster::clamp(imgs.mod.raw, lower = -1, upper = 1)
\end{Sinput}
\end{Schunk}

\pkg{RGISTools} includes the function \fct{genPlotGIS} to display satellite 
imagery. \fct{genPlotGIS} is a wrapper function of \pkg{tmap} \citep{tmap:2018}
with options and layers configured to easily display the spatial information
dealt within \pkg{RGISTools}:

\begin{Schunk}
\begin{Sinput}
R> genPlotGIS(r = imgs.mod.ndvi,  
+             region = ex.navarre,
+             zlim = c(0,1),
+             tm.raster.r.palette = rev(terrain.colors(40)),
+             tm.graticules.labels.size = 1.3,
+             tm.graticules.n.x = 2,
+             tm.graticules.n.y = 2,
+             tm.graticules.labels.rot = c(0,90),
+             panel.label.size = 1.5)
\end{Sinput}
\end{Schunk}

\subsection{Processing satellite imagery} \label{subsec:process}

Processing comprises cloud masking, filling data gaps, and smoothing outliers
from the imagery. Cloud masking and filling gaps are straightforward through
image compositing (\fct{genCompositions}). This technique combines several 
images within sequential time windows into a single image, by selecting or
smoothing the values per pixel over time. Compositing improves the quality of
the images but it also reduces the amount of information available. Less
information may lead to a lower accuracy in subsequent analyses
\citep{compositing:2011}.

\pkg{RGISTools} offers an alternative to preserve as much data as possible.
Cloudy pixels can be masked using the quality bands of optical multispectral
images. Then, data gaps can be filled and outliers smoothed with a statistical
technique called the interpolation of the mean anomalies (IMA)
\citep{imaSmooth:2018, imaCloud:2019}. Since the latter is more sophisticated,
we elaborate on this alternative in the following paragraphs.

\subsubsection{Cloud masking} \label{subsubsec:cloud}

Satellite programs apply their own methodology to determine the pixels covered
by clouds \citep{cloud:2018}. The results are saved in the quality bands of 
level-2 products, together with other information affecting the quality of the
surface reflectance estimates \citep*[see e.g.,][]{mod09ug}. The functions
\fct{lsCloudMask}, \fct{modCloudMask}, and \fct{senCloudMask} interpret the
quality bands in each program and save time-series of cloud masks to disk.
In these masks, clear-sky and cloudy pixels are represented by \verb|1|s and
\verb|NA|s respectively.

The following code extracts the cloud masks for the MODIS time-series. The
masks are placed by \fct{modCloudMask} in a new folder defined by 
\code{out.name} in the \code{wdir} directory (i.e., \verb|./Modis/mod_cldmask|):

\begin{Schunk}
\begin{Sinput}
R> modCloudMask(src = wdir.mod.mosaic,
+               out.name = "mod_cldmask",
+               AppRoot = wdir.mod)
\end{Sinput}
\end{Schunk}

Masks are saved as GTiff files, which can be imported into \proglang{R}.
In the following chunk of code, the files with the cloud masks are listed and
loaded as a \class{stack}. As cloud masks contain categorical values, they
must be converted into \class{factor} with the function \fct{ratify}:

\begin{Schunk}
\begin{Sinput}
R> wdir.mod.cld <- file.path(wdir.mod, "mod_cldmask")
R> files.mod.cld <- list.files(wdir.mod.cld, full.names = TRUE)
R> imgs.mod.cld <- raster::stack(files.mod.cld)
R> imgs.mod.cld <- raster::stack(lapply(as.list(imgs.mod.cld), ratify))
\end{Sinput}
\end{Schunk}

Cloud masks could be on a coarser scale (here, $1 \times 1 \, km^{2}$) than
the multispectral images ($0.5 \times 0.5 \, km^{2}$). Masks can be resampled
with the \fct{projectRaster} function to obtain rasters at the same resolution
as the multispectral images. Since the masks are categorical values ($1$s for
clear-sky and \verb|NA|s for cloudy pixels), the resampling is carried out with
the nearest neighbor method. Cloud masks can be applied to the NDVI images as
follows:

\begin{Schunk}
\begin{Sinput}
R> imgs.mod.masks <- raster::projectRaster(imgs.mod.cld,
+                                          imgs.mod.ndvi[[1]],
+                                          method = "ngb")
R> imgs.mod.ndvimks <- imgs.mod.masks * imgs.mod.ndvi
R> names(imgs.mod.ndvimks) <- names(imgs.mod.ndvi)
\end{Sinput}
\end{Schunk}

\subsubsection{Gap-filling and smoothing} \label{subsubsec:fillsmooth}

Cloud removal or sensor failures can lead to data gaps in the time-series
of satellite images. Additionally, noise from aerosols, dust, and sensor 
measurement errors can reduce the quality of the remotely sensed data. 
Many gap-filling and smoothing approaches have been developed to mitigate
these issues \citep{missing:2015}. Among them, there is the IMA procedure,
which was developed by \cite{imaSmooth:2018, imaCloud:2019}. \pkg{RGISTools} 
implements a generic version of the algorithm in the \fct{genSmoothingIMA}
and \fct{genSmoothingCovIMA} functions.

IMA borrows information from a temporal neighborhood of the image to be filled
or smoothed (target image henceforth). The neighborhood extends around the
images that are assumed to be similar to the target image. Two parameters
confine the size of the neighborhood; \code{nDays}, that is, the number of days
before and after the capturing date of the target image, and \code{nYears},
which is the number of previous and subsequent years. For instance, if
\code{nDays = 1} and \code{nYears = 1}, the neighborhood is built from images
within a period of 1 day before and after the target image plus images from
the same days of the year but in the previous and subsequent years. IMA uses
incomplete neighborhoods in case some images do not exist. Then, the function
conducts the following steps:

\begin{enumerate}
  \item Obtain the average image of the neighboring images.
  \item Subtract the average image from the target image to obtain an image of
  anomalies.
  \item Screen out the anomalies outside a range of percentiles (e.g., 
  0.05-0.95).
  \item Aggregate the anomaly image into a coarser resolution using the mean or
  median (\code{fun} argument) and an aggregation factor (\code{fact} argument).
  For instance, \code{fun = 'mean'} and \code{fact = 4} averages sets of 4 
  pixels into a single pixel.
  \item Interpolate the aggregated image of anomalies using thin-plate splines
  from the \pkg{fields} package \citep{fields:2017}.
  \item Predict the target image in the original resolution adding the
  interpolated anomalies and the average image.
\end{enumerate}

The size of the neighborhood, the aggregation factor, and the range of 
percentiles should be adapted in each situation to get the best performance
from IMA. For instance, the \code{nDays} should be adjusted based on the 
temporal resolution of the of the time-series of images. Also, cloudy
series may require larger neighborhoods. We recommend that the neighborhood
extends over days rather than years, when there is little resemblance between
seasons. Finally, narrower percentiles might be considered when handling more
pre-processed data products.

The \fct{genSmoothingIMA} and \fct{genSmoothingCovIMA} functions take as an
input a time-series of satellite images in the form of a \class{RasterStack}
(\code{rStack} argument) with their capturing dates included in the names of
the layers as \verb|YYYYJJJ| (\verb|Y| and \verb|J| represent a year and
Julian day digits). This format happens naturally when the user follows the
workflow in \pkg{RGISTools} (see the code below). The \code{Img2Fill} argument
sets which are the target images of the \code{rStack}.

The difference between \fct{genSmoothingIMA} and \fct{genSmoothingCovIMA} lies
in the use of covariates in step 5. A \class{RasterStack} of covariates can
specified with the argument \code{cStack}, which must have the same dimensions
as the \code{rStack}.

IMA functions return a \class{stack} that only fills the missing values and
preserves the original target image if \code{only.na = TRUE}. By default, the
option equals to \code{FALSE}, so the functions return entirely predicted 
target image.

\begin{figure}[t!]
\centering
\includegraphics{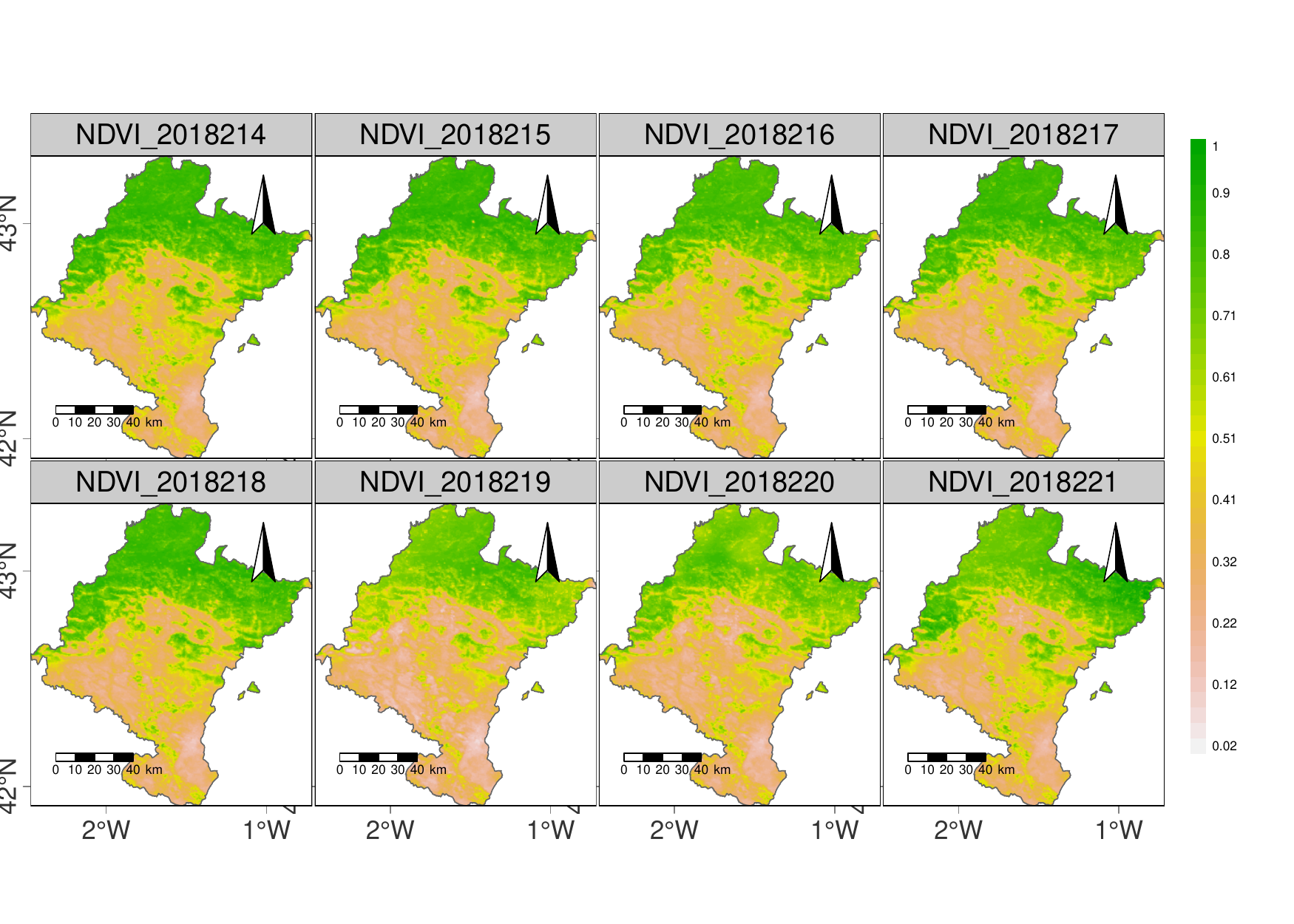}
\vspace*{-10mm}
\caption{\label{fig:imafig} Reconstructed NDVI from cloud-masked images
using the interpolation of the mean anomalies (IMA) procedure. The scenes
cover August $2^{nd}-9^{th}$, 2019 (2018220-2018221 in \texttt{YYYYJJJ}
format).}
\end{figure}

The following code fills the empty pixels of the entire series of satellite
images (Figure~\ref{fig:imafig}). The blank spaces caused by the cloud masks
are filled by the IMA procedure (Figure~\ref{fig:imafig}) using a neighborhood
of 8 days from the same year of the target image. IMA does not guarantee that
the prediction of the NDVI stays in the [-1,1] range, so the results must be
truncated with the \fct{clamp} function from \pkg{raster}. To look at the
dataset for our ROI alone, we mask the pixels outside Navarre with \fct{mask}:

\begin{Schunk}
\begin{Sinput}
R> imgs.mod.imaraw <- genSmoothingIMA(rStack = imgs.mod.ndvimks,
+                                     Img2Fill = 1:nlayers(imgs.mod.ndvimks),
+                                     nDays = 8, 
+                                     nYears = 1,
+                                     aFilter = c(0.05, 0.95),
+                                     fact = 8)
R> imgs.mod.imaclamp <- raster::clamp(imgs.mod.imaraw, lower = -1, upper = 1)
R> ex.mod.navarre <- sf::st_transform(ex.navarre, 
+                                     crs = projection(imgs.mod.imaclamp))
R> imgs.mod.imanavarre <- raster::mask(imgs.mod.imaclamp, ex.mod.navarre)
R> genPlotGIS(imgs.mod.imanavarre,
+             region = ex.mod.navarre,
+             tm.graticules.labels.size = 1.3,
+             tm.graticules.n.x = 2,
+             tm.graticules.n.y = 2,
+             tm.graticules.labels.rot = c(0,90),
+             panel.label.size = 1.5,
+             tm.raster.r.palette = rev(terrain.colors(40)))
\end{Sinput}
\end{Schunk}

IMA can be used with datasets retrieved or loaded with other packages. Other
classes, such as \class{stars} or \class{satellite} objects, can be easily 
coerced into \class{RasterStack}. To facilitate the interoperability of IMA
with other packages, the function allows to pass the capturing dates of the 
imagery as a vector of \class{Dates} class objects through the argument
\code{r.dates}.

\section{Working example} \label{sec:example}

In this section, we present a case study that combines Landsat-8 and Sentinel-2
imagery to monitor the water level of a reservoir in Northern Spain.
Section~\ref{subsec:egroi} defines the ROI and introduces the auxiliary data
required for this exercise (topographic data and water level observations).
Section~\ref{subsec:egretrieve} retrieves Landsat-8 and Sentinel-2 images for
the period and the region of analysis. In Section~\ref{subsec:egcustom}, the
satellite imagery is customized (cropping and computing a remote sensing index)
to detect the surface of the water body. Section~\ref{subsec:detection}
translates the flooded area into water levels with the aid of the topographic 
map. Finally, results are contrasted with the \emph{in situ} measurements.

The working example takes $81.24$ GB of memory space and the overall running
time is less than $3$ hours. However, it is divided into shorter parts, whose
results are available via downloadable files. Thus, the code in each part can be
reproduced independently from each other. The demand of time and memory space
decreases throughout the example, being the maximum $80.5$ GB and $2.2$ hours
to run Section~\ref{subsec:egretrieve} and the minimum $0.07$ GB and nearly $3$
seconds to complete Section~\ref{subsec:egcustom}.

\subsection{Region of interest} \label{subsec:egroi}

We examine the Itoiz reservoir, which is located in Northern Spain within the
region of Navarre. The dam was built to collect the waters from the Irati river.
The reservoir is located northeast the village of Aoiz, in the foothills of
the Pyrenees. The pond extends over $1100$ ha and has a capacity of $418$
$hm^{3}$. The reservoir became fully operational in 2006.

In the following code, the spatial domain of the water body is defined using
the \pkg{sf} package \citep{sf:2018}. The area is delimited by a \class{bbox}
with the minimum and maximum longitude-latitude coordinates. The \class{bbox}
is transformed into a \class{sfc} class object to create a rectangular polygon,
and then turned into an \class{sf} object:

\begin{Schunk}
\begin{Sinput}
R> roi.bbox <- sf::st_bbox(c(xmin = -1.40,
+                            xmax = -1.30,
+                            ymin = 42.79,
+                            ymax = 42.88),
+                      crs = 4326)
R> roi.sfc <- sf::st_as_sfc(roi.bbox)
R> roi.sf <- sf::st_as_sf(roi.sfc)
\end{Sinput}
\end{Schunk}

\begin{figure}[t!]
\centering
\includegraphics{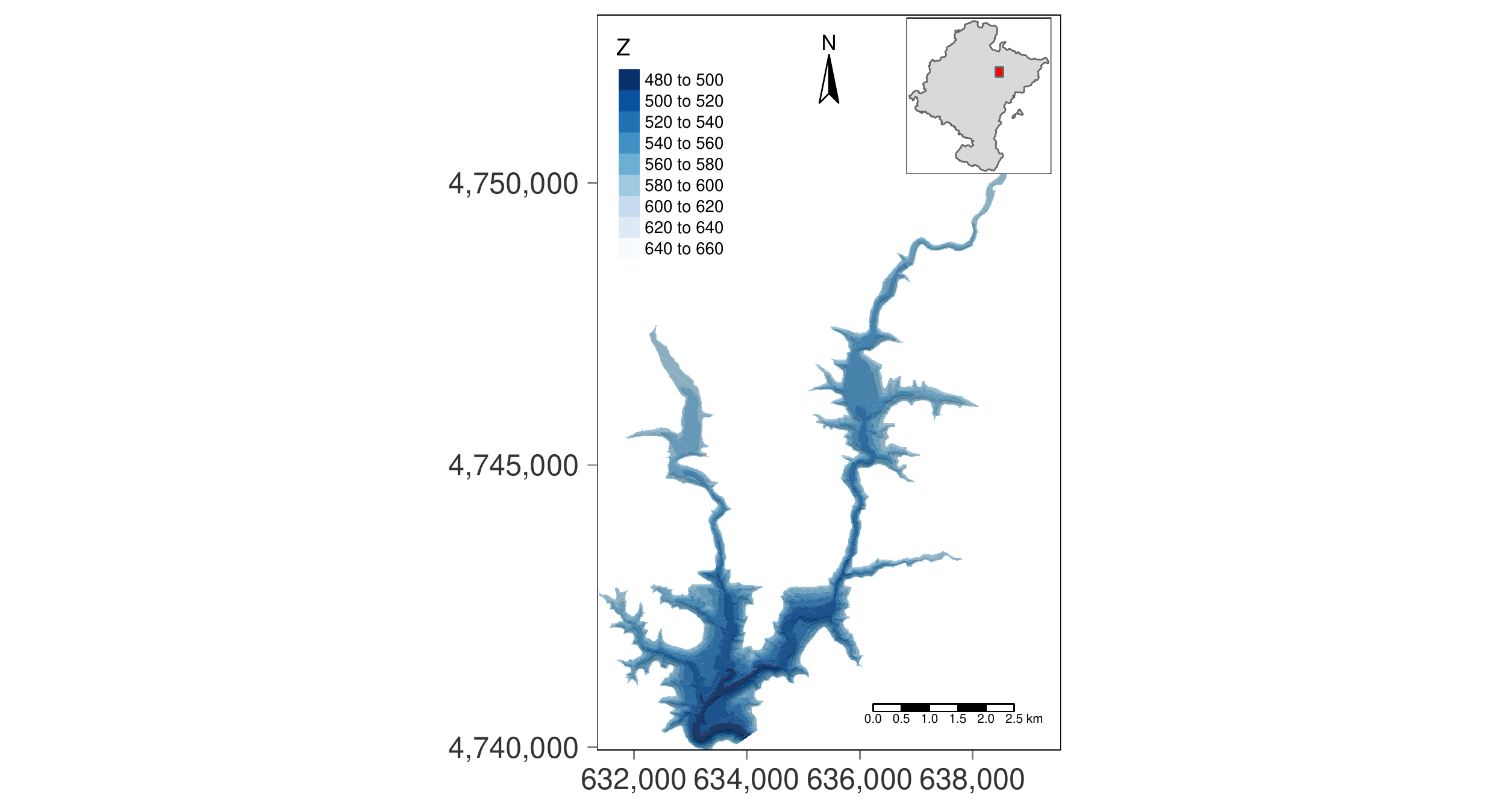}
\caption{\label{fig:topofig} Elevation map of the basin of the Itoiz reservoir.
The elevation (\texttt{Z}) is measured in meters above sea level (m.a.s.l.).
The map was derived from freely available information provided via online by
the local administration \citep{idena:2019}.}
\end{figure}

The water level refers to the elevation reached by the pond's shoreline, which
can be derived by superimposing the flooded area and a topographic map. A
contour map is freely available at the local government's website
\citep{idena:2019}, which was interpolated to a 10 meter resolution map 
applying the inverse distance weighting (IDW) method from \pkg{gstat}
\citep{gstat:2004}. The elevation map (Figure~\ref{fig:topofig}) was named as
\code{altimetry.itoiz} and saved as a \class{RasterLayer} into an ``RData''
file.

The map ($0.77$ MB) is available at the link provided in the reference
\cite{aux:2019:itoiz}. The file should be downloaded, unzipped, and placed in
the \code{wdir} directory. Then, the map can be loaded as:

\begin{Schunk}
\begin{Sinput}
R> wdir.topo <- file.path(wdir, "aux_info", "topography_Itoiz.RData")
R> load(wdir.topo)
\end{Sinput}
\end{Schunk}

As mentioned earlier, the estimates will be compared with \emph{in situ}
observations. Water levels are measured on a daily basis at the dam wall and
made publicly available at the Automatic Hydrological Information System of the
Ebro River Basin Authority \citep{saih:2019}. The file is available at the 
reference provided above and can be loaded as follows:

\begin{Schunk}
\begin{Sinput}
R> wdir.levels <- file.path(wdir, "aux_info", "level_itoiz.csv")
R> obs.itoiz <- read.csv(wdir.levels, colClasses = c("Date", "numeric"))
\end{Sinput}
\end{Schunk}

\subsection{Retrieving satellite imagery} \label{subsec:egretrieve}

\subsubsection{Finding a time-series} \label{subsubsec:egsearch}

The functions \fct{lsSearch} and \fct{senSearch} scan the Landsat and Sentinel-2
repositories to find those scenes that match the requested data product 
(\code{product}), time interval (\code{dates}), and ROI (\code{region = roi.sf}).
In this working example, we want to track the water levels from mid summer 2018
to mid spring 2019 (i.e., \code{dates = as.Date("2018-07-01") + seq(0, 304, 1)}),
as this is the time of the season that water storage varies the most.

Landsat and Sentinel search functions allow to filter the results by cloud
coverage. Discarding cloudy images at an early stage can save space in the
disk and processing time. The cloud coverage filter can be set with the 
\code{cloudCover} argument, indicating the lower and upper percentages of 
the pixels of an image being covered by clouds. The view of the reservoir is
likely obstructed by clouds during winter, since it is located in a mountainous 
area. Hence, we restrict our search to images with a cloud coverage below $80\%$
(\code{cloudCover = c(0,80)}).

We use the surface reflectance product to perform our analysis, i.e., imagery
that has been atmospherically corrected (level-2). Landsat only provides
immediate access to level-1 products (\code{product = "LANDSAT_8_C1"}), so in
order to obtain the level-2 product, we must search level-1 images first and
then, at the time of downloading, request their correction to the Earth
Resources Observation and Science (EROS) Center through their Science 
Processing Architecture (ESPA) \citep{espa2019}:

\begin{Schunk}
\begin{Sinput}
R> library("RGISTools")
R> sres.ls8 <- lsSearch(product = "LANDSAT_8_C1",
+                       dates = as.Date("2018-07-01") + seq(0, 304, 1),
+                       region = roi.sf,
+                       cloudCover = c(0,80),
+                       username = "<USERNAME>",
+                       password = "<PASSWORD>")
\end{Sinput}
\end{Schunk}

The function \fct{lsSearch} returns a \class{data.frame} with the images that
were found as rows and their metadata details as columns. Regarding Sentinel,
surface reflectance images are available from the Sentinel-2 mission with the
product ``S2MSI2A'' (Sentinel-2 MultiSpectral level-2A):

\begin{Schunk}
\begin{Sinput}
R> sres.sn2 <- senSearch(platform = "Sentinel-2",
+                        product = "S2MSI2A",
+                        dates = as.Date("2018-07-01") + seq(0, 304, 1),
+                        region = roi.sf,
+                        cloudCover = c(0,80),
+                        username = "<USERNAME>",
+                        password = "<PASSWORD>")
\end{Sinput}
\end{Schunk}

Note that both \fct{lsSearch} and \fct{senSearch} require the log-in 
credentials in contrast to \fct{modSearch}. The credentials are required to
access the information available at EarthExplorer and SciHub. Replace the
\verb|<USERNAME>| and \verb|<PASSWORD>| with your own credentials after signing
up for both web services \citep{earthdata:reg, scihub:reg}. The \fct{senSearch}
function returns a vector of URLs.

\subsubsection{Downloading} \label{subsubsec:egdownload}

The \fct{lsDownload} and \fct{senDownload} functions retrieve the time-series
of satellite images found in the previous section (\code{sres.ls8} and 
\code{sres.sn2}). Be aware that downloading satellite images can be 
time-consuming and requires enough storage space in the disk ($2.2$ hours and
$80.5$ GB). In case of insufficient memory space, you can skip this section and
download the results concerning Landsat-8 ($7.66$ GB) \citep{ls8:2019:dw} and
Sentinel-2 ($12$ GB) images\citep{sn2:2019:dw} or get the results from
subsequent milestones.

As mentioned earlier, Landsat-8 images must be atmospherically corrected by EROS.
By setting \code{lvl = 2}, \fct{lsDownload} makes a request to ESPA to process
the list of level-1 images (\code{sres.ls8}) and gets the corresponding level-2
from their response. To distinguish this request from previous ones, the 
petition should be named using the \code{l2rqname} argument. The downloaded 
files are directly saved in the \verb|./Landsat8/raw| directory. For our 
purpose, we only require the green (``band3'') and near infra-red (``band5'')
bands from the multispectral images to compute the NDWI \citep{ndwi:1996} and
the quality (``pixel\_qa'') band to analyse the cloud coverage. The 
\code{bFilter} argument allows to extract specific bands, which are then saved
as GTiffs in the \verb|./Landsat8/untar| directory. Once the transformation is
completed, the original files could be removed to free up memory space by
adding \code{raw.rm = TRUE}:

\begin{Schunk}
\begin{Sinput}
R> wdir.ls8 <- file.path(wdir, "Landsat8")
\end{Sinput}
\end{Schunk}
\begin{Schunk}
\begin{Sinput}
R> lsDownload(searchres = sres.ls8,
+             lvl = 2,
+             untar = TRUE,
+             bFilter = list("band3", "band5", "pixel_qa"),
+             username = "<USERNAME>",
+             password = "<PASSWORD>",
+             l2rqname = "<REQUESTNAME>",
+             raw.rm = TRUE,
+             AppRoot = wdir)
\end{Sinput}
\end{Schunk}

Similarly, \fct{senDownload} downloads and uncompressed images from Sentinel.
In Sentinel-2, the bands 3 and 8 correspond to green a near infra-red wavelengths.
Both bands are available at a maximum resolution of $10$$m$, so we refer to them
as ``B03\_10m'' and ``B08\_10m''. The cloud coverage is captured by the cloud
probability band (CLDPRB), which is available at a maximum resolution of $20$
$m$ (``CLDPRB\_20m''). In the code that follows, the function downloads the
file in \verb|./Sentinel2/raw| directory, extracts the bands, and saves them
in the \verb|./Sentinel2/unzip| directory. To clear memory space, we specify
\code{raw.rm = TRUE} to delete the original files in \verb|./Sentinel2/raw|:

\begin{Schunk}
\begin{Sinput}
R> wdir.sn2 <- file.path(wdir, "Sentinel2")
\end{Sinput}
\end{Schunk}
\begin{Schunk}
\begin{Sinput}
R> senDownload(searchres = sres.sn2,
+              unzip = TRUE,
+              bFilter = list("B03_10m", "B08_10m", "CLDPRB_20m"),
+              username = "<USERNAME>",
+              password = "<PASSWORD>",
+              raw.rm = TRUE,
+              AppRoot = wdir.sn2)
\end{Sinput}
\end{Schunk}

\subsection{Customizing satellite imagery} \label{subsec:egcustom}

\subsubsection{Mosaicking and cropping} \label{subsubsec:egcrop}

Due to the size of the ROI, it is computationally and memory efficient to remove
the pixels outside \code{roi.sf}. The next code applies \fct{lsMosaic} and
\fct{senMosaic} to the images saved in \verb|./Landsat8/untar| and 
\verb|./Sentinel2/unzip|. The results are placed in two folders created
automatically by the \verb|Mosaic| functions; \verb|./Landsat8/ls8_itoiz| and
\verb|./Sentinel2/sn2_itoiz|):

\begin{Schunk}
\begin{Sinput}
R> wdir.ls8.untar <- file.path(wdir.ls8, "untar")
R> lsMosaic(src = wdir.ls8.untar,
+           region = roi.sf,
+           out.name = "ls8_itoiz",
+           gutils = TRUE,
+           AppRoot = wdir.ls8)
R> wdir.sn2.unzip <- file.path(wdir.sn2, "unzip")
R> senMosaic(src = wdir.sn2.unzip,
+            region = roi.sf,
+            out.name = "sn2_itoiz",
+            gutils = TRUE,
+            AppRoot = wdir.sn2)
\end{Sinput}
\end{Schunk}

The original tiles are not required for the subsequent steps of the analysis,
so we remove them to clear memory space as follows:

\begin{Schunk}
\begin{Sinput}
R> unlink(wdir.ls8.untar, recursive = TRUE)
R> unlink(wdir.sn2.unzip, recursive = TRUE)
\end{Sinput}
\end{Schunk}

Cropping the series of images requires roughly $15$ minutes and the results
occupy $280$ MB of memory in the hard disk. If needed, the results are 
available at \cite{ls8:2019:mos} for Landsat-8 and at \cite{sn2:2019:mos} for
Sentinel-2. The following steps require the files to be uncompressed and saved
in two folders called \verb|./Landsat8| and \verb|./Sentinel2| in the \code{wdir}
directory.

\subsubsection{Cloud mask filtering} \label{subsubsec:egcladmask}

Clouds in the area may hamper the identification of the water-body shoreline. 
Here, we inspect the cloudiness at the reservoir by extracting and analyzing the
cloud masks. The \fct{lsCloudMask} and \fct{senCloudMask} functions interpret
the information about the presence of clouds from the quality bands. The
location of these quality bands must be indicated through the \code{src}
argument. The generated cloud masks are saved in \code{AppRoot} directory,
in a new folder named as the \code{out.name} argument:

\begin{Schunk}
\begin{Sinput}
R> wdir.ls8.mosaic <- file.path(wdir.ls8, "ls8_itoiz")
R> lsCloudMask(src = wdir.ls8.mosaic,
+              out.name = "ls8_cldmask",
+              AppRoot = wdir.ls8)
R> wdir.sn2.mosaic <- file.path(wdir.sn2, "sn2_itoiz")
R> senCloudMask(src = wdir.sn2.mosaic,
+               out.name = "sn2_cldmask",
+               AppRoot = wdir.sn2)
\end{Sinput}
\end{Schunk}

The quality bands are translated into cloud masks in few seconds for both
series of images and the outputs take $2.14$ MB of memory. Results are 
available in \cite{ls8:2019:cld} and \citep{sn2:2019:cld} for Landsat-8 and
Sentinel-2 respectively. Download the files and unzip them at \verb|./Landsat8|
and \verb|./Sentinel2|.

In the following, we load the cloud masks to conduct further analyses:

\begin{Schunk}
\begin{Sinput}
R> wdir.ls8.cld <- file.path(wdir.ls8, "ls8_cldmask")
R> wdir.sn2.cld <- file.path(wdir.sn2, "sn2_cldmask")
R> wdir.all.cld <- list(wdir.ls8.cld, wdir.sn2.cld)
R> files.cld.msk <- lapply(wdir.all.cld, list.files, full.names = TRUE)
R> imgs.cld.msk <- lapply(files.cld.msk, raster::stack)
R> names(imgs.cld.msk) <- c("ls8", "sn2")
\end{Sinput}
\end{Schunk}

The next code finds the dates in which the cloud coverage remained below a
threshold at the Itoiz reservoir. The threshold was set to $30\%$ for Landsat-8
and $0.1\%$ for Sentinel-2 images. These thresholds were decided through visual
inspection of the images and the cloud masks. Landsat-8 has a higher threshold
than Sentinel-2 due to missclassified water pixels as clouds by the Landsat-8 
algorithms:

\begin{Schunk}
\begin{Sinput}
R> cld.coverage <- lapply(imgs.cld.msk,
+                         function(x){colSums(is.na(getValues(x)))/ncell(x)})
R> names(cld.coverage) <- c("ls8", "sn2")
R> ls8.clr.dates <- genGetDates(names(imgs.cld.msk$ls8))[cld.coverage$ls8 < 0.30]
R> sn2.clr.dates <- genGetDates(names(imgs.cld.msk$sn2))[cld.coverage$sn2 < 0.001]
\end{Sinput}
\end{Schunk}

Both \code{ls8.clr.dates} and \code{sn2.clr.dates} represent the dates with
clear skies at the reservoir.

\subsubsection{Computing the NDWI} \label{subsubsec:egndwi}

The Normalized Difference Water Index (NDWI) is a remote sensing index usually
applied for detecting flooded areas \citep{ndwi:1996}. It has been used extensively
to map water bodies from multispectral satellite images \citep{ndwi:2016}.
The NDWI marks out water bodies based on the strong absorbability in the near
infra-red band (NIR) and the strong reflectance in the green band from water. 
Pixels with a NDWI above 0 are candidates for open water bodies, although
thresholds between 0 and 0.1 are frequently adopted \citep{ndwi:2009}.

\pkg{RGISTools} provides a built-in function to compute the NDWI (\fct{varNDWI}). 
In the following block of code, both \fct{ls8FolderToVar} and \fct{senFolderToVar}
apply \fct{varNDWI} to the time-series of images considered so far. Note that
both \verb|FolderToVar| functions use the same definition of the NDWI, in spite
of the discrepancies between the band names and numbers of the Landsat-8 and
Sentinel-2 missions. The functions \verb|FolderToVar| are responsible for
matching the band names in \fct{varNDWI} with the appropriate band numbers in
each mission. The NDVI is only computed for the clear-sky dates, which were
obtained in the previous section:

\begin{Schunk}
\begin{Sinput}
R> wdir.ls8.mosaic <- file.path(wdir.ls8, "ls8_itoiz")
R> ls8FolderToVar(src = wdir.ls8.mosaic,
+                 fun = varNDWI,
+                 dates = ls8.clr.dates,
+                 AppRoot = wdir.ls8)
R> wdir.sn2.mosaic <- file.path(wdir.sn2, "sn2_itoiz")
R> senFolderToVar(src = wdir.sn2.mosaic,
+                 fun = varNDWI,
+                 dates = sn2.clr.dates,
+                 AppRoot = wdir.sn2)
\end{Sinput}
\end{Schunk}

The time-series of NDWIs from the Landsat-8 and Sentinel-2 imagery are 
automatically saved at \verb|./Landsat8/NDWI| and \verb|./Sentinel2/NDWI|
respectively. The overall computational time is a few minutes for both series 
and the NDWI imagery occupies $73.22$ MB of space.

The files are available at \cite{ls8:2019:ndwi} and \cite{sn2:2019:ndwi}.
Once downloaded, unzip the files and save them at \verb|./Landsat8| and
\verb|./Sentinel2| in the \code{wdir} directory. Henceforth, no more
dowloadable files will be provided.

\subsection{Detecting water and water level analysis} \label{subsec:detection}

The NDWI images can be loaded in \proglang{R} using the \fct{stack} function
from the \pkg{raster} package. Images from Landsat-8 and Sentinel-2 must be
loaded separately since their resolutions is different (30 and 10 meters,
respectively). The \fct{stack} function returns a \class{RasterStack} where
each layer is an NDWI image of the time-series:

\begin{Schunk}
\begin{Sinput}
R> imgs.ndwi <- list(
+            stack(list.files(file.path(wdir.ls8,"NDWI"), full.names = TRUE)),
+            stack(list.files(file.path(wdir.sn2,"NDWI"), full.names = TRUE)))
\end{Sinput}
\end{Schunk}

Layers receive the name of the index and their capturing date (e.g.,
``NDWI\_2018244''). To keep track of the source of every image, we additionally
paste a platform label (``LS8'' and ``SN2'') to the names of the layers: 

\begin{Schunk}
\begin{Sinput}
R> names(imgs.ndwi[[1]]) <- paste0(names(imgs.ndwi[[1]]), "_LS8")
R> names(imgs.ndwi[[2]]) <- paste0(gsub("10m", "SN2", names(imgs.ndwi[[2]])))
\end{Sinput}
\end{Schunk}

The following code combines the Landsat-8 and Sentinel-2 time-series into a
single \class{stack} to simplify the analysis. The function \fct{projectRaster}
modifies the coordinate reference system and the resolution from the Sentinel-2
imagery to match those in the Landsat-8 series. Both are combined into a single
\class{stack}  as follows: 

\begin{Schunk}
\begin{Sinput}
R> imgs.ndwi[[2]] <- raster::projectRaster(imgs.ndwi[[2]], imgs.ndwi[[1]])
R> imgs.ndwi <- raster::stack(imgs.ndwi)
\end{Sinput}
\end{Schunk}

Then, the layers are rearranged to follow the temporal sequence:

\begin{Schunk}
\begin{Sinput}
R> imgs.ndwi <- imgs.ndwi[[order(names(imgs.ndwi))]]
\end{Sinput}
\end{Schunk}

\begin{figure}[t!]
\centering
\includegraphics{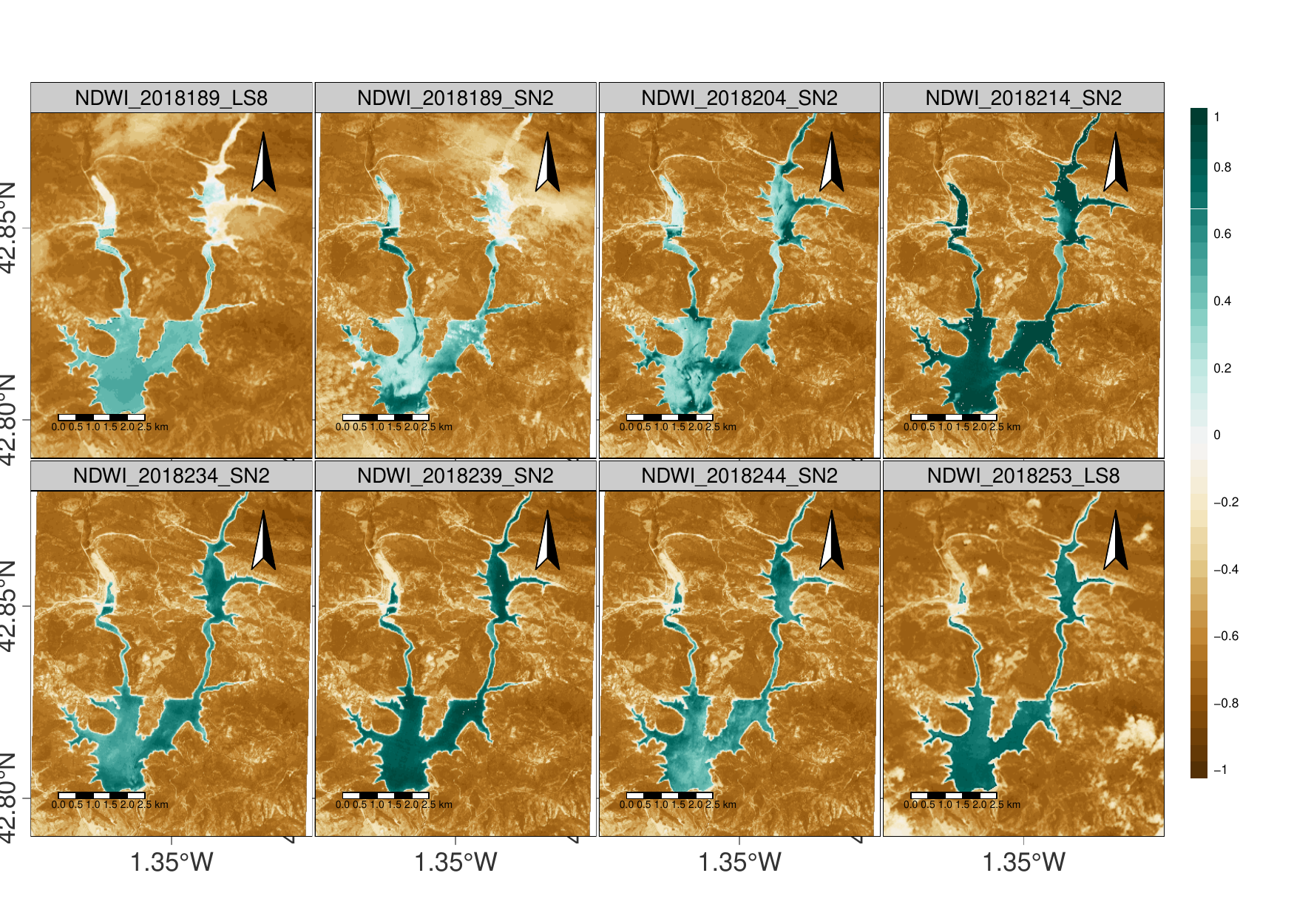}
\vspace*{0mm}
\caption{\label{fig:ndwifig} Water detection (green color) at the Itoiz 
reservoir. The first $8$ instances in the time-series of images of NDWI from
Landsat-8 (abbreviated as ``LS8'') and Sentinel-2 (``SN2''). The ``x'' and ``y''
axes are the longitude and latitude coordinates. The names of the panels
additionally show the  capturing date of the image in \texttt{YYYYJJJ} format,
where \texttt{Y} represents a year digit and \texttt{J} is a Julian date digit.}
\end{figure}

We inspect the results showing the first $8$ images in \code{imgs.ndwi}
(Figure~\ref{fig:ndwifig}):

\begin{Schunk}
\begin{Sinput}
R> genPlotGIS(imgs.ndwi[[1:8]],
+             zlim = c(-1,1),
+             tm.raster.r.palette = "BrBG",
+             tm.graticules.labels.size = 1.3,
+             tm.graticules.n.x = 2,
+             tm.graticules.n.y = 2,
+             tm.graticules.labels.rot = c(0,90),
+             panel.label.size = 1)
\end{Sinput}
\end{Schunk}

For consistency, the elevation map is also projected to match the reference
system of the NDWI dataset:

\begin{Schunk}
\begin{Sinput}
R> map.z <- raster::projectRaster(altimetry.itoiz,
+                                 crs = st_crs(imgs.ndwi)$proj4string,
+                                 method = "bilinear")
\end{Sinput}
\end{Schunk}

Translating the NDWI into water levels takes place as follows:
\begin{enumerate}
  \item Cells representing flooded areas are converted into polygons. Here,
  pixels above $-0.1$ (selected by visual inspection) are considered as flooded
  and converted into polygons with \fct{rasterToPolygons}.
  \item The boundaries of neighboring cells are resolved, and just the edges of
  the water bodies remain after \fct{st\_union}. The output is a
  \class{MULTIPOLYGON}, which is then coerced into separate \class{POLYGON}s
  by \fct{st\_cast}.
  \item The main water body is distinguished from auxiliary reservoirs and
  isolated missclassified pixels by finding the polygon with maximum area.
  The function \fct{st\_area} computes the area for each polygon.
  \item The elevation map is masked with the line-strings of the shoreline of
  the main water body, which removes every elevation pixel outside the 
  trajectory of the borderline.
  \item The median of the shoreline's elevation gives the estimated water level
  \code{level.est}. The median allows to better counteract errors due to the
  interpolation of the topographic map and the detection of the shoreline due
  to the pixel resolution.
\end{enumerate}

\begin{Schunk}
\begin{Sinput}
R> shorelns <- lapply(as.list(imgs.ndwi),
+                     function(r){
+                       water <- raster::rasterToPolygons(clump(r> -0.1),
+                                                         dissolve = TRUE)
+                       shores <- sf::st_union(sf::st_as_sfc(water))
+                       bodies <- sf::st_cast(shores, "POLYGON")
+                       areas  <- sf::st_area(bodies)
+                       sf::st_sf(
+                         sf::st_cast(
+                           bodies[which(areas == max(areas))],
+                           "MULTILINESTRING"))})
R> shorelns.z <- raster::stack(lapply(shorelns,
+                                     function(x, map.z){
+                                       mask(map.z, x)},
+                                     map.z))
R> level.est <- cellStats(shorelns.z, 'median')
\end{Sinput}
\end{Schunk}

\begin{figure}[t!]
\centering
\includegraphics{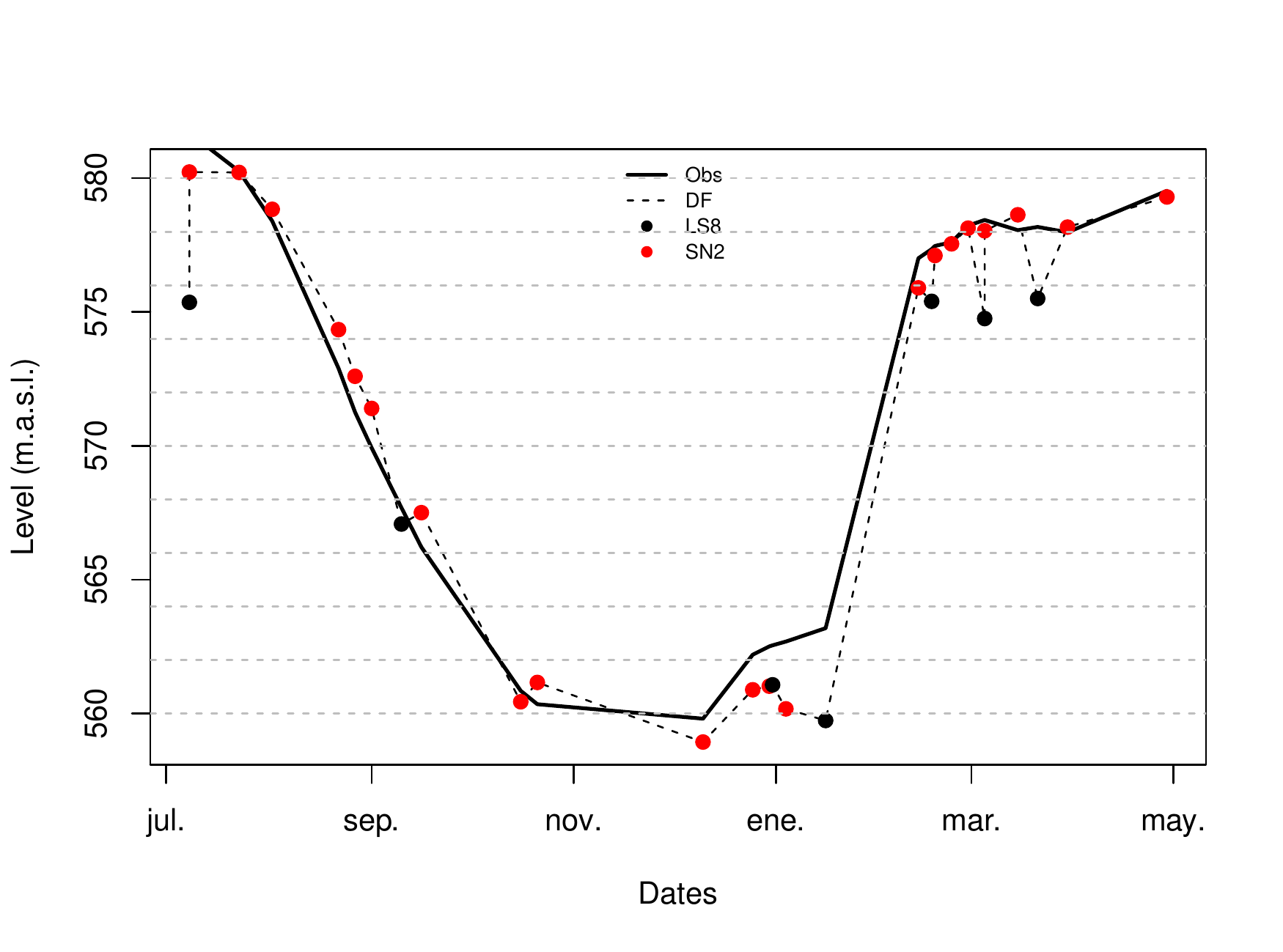}
\vspace*{0mm}
\caption{\label{fig:egresults} Water levels in the Itoiz reservoir between
August 2018 and May 2019. The water levels are in meters above sea level
(m.a.s.l.). The black line represents the observations. Black and red 
dots are estimates from Landsat-8 and Sentinel-2 respectively. The dashed
line represents the combination of Landsat-8 and Sentinel-2 water levels.}
\end{figure}

To sum up, we build a \class{data.frame} where the rows represent the sequence
of images in the time-series and the columns represent key aspects of the
analysis such as, the satellite mission (\verb|sat|), the capturing date of
the image (\verb|date|), the observed water levels (\verb|obs|), and the
estimated water level (\verb|est|). This \class{data.frame} summarizes 
the results of the case study (Figure~\ref{fig:egresults}):

\begin{Schunk}
\begin{Sinput}
R> no.imgs <- nlayers(imgs.ndwi)
R> results <- data.frame("sat" = character(no.imgs),
+                        "date" = structure(integer(no.imgs), class = "Date"),
+                        "obs" = double(no.imgs),
+                        "est" = double(no.imgs),
+                        stringsAsFactors = FALSE)
R> results$sat <- ifelse(grepl("LS8", names(imgs.ndwi)), "LS8", "SN2")
R> results$date <- genGetDates(names(imgs.ndwi))
R> results$obs <-  merge(obs.itoiz,results)$level.masl
R> results$est <- level.est
\end{Sinput}
\end{Schunk}

Figure~\ref{fig:egresults} shows that the measured water levels are closely
followed by the estimates, especially by Sentinel-2. Figure~\ref{fig:egresults}
also shows how Landsat-8 and Sentinel-2 complement each other to gain temporal
coverage. We finally compute some metrics of the performance:

\begin{Schunk}
\begin{Sinput}
R> error <- results$obs - results$est
R> mean(abs(error), na.rm = TRUE)
\end{Sinput}
\begin{Soutput}
[1] 1.35971
\end{Soutput}
\begin{Sinput}
R> mean(abs(error)[results[,"sat"] == "LS8"], na.rm = TRUE)
\end{Sinput}
\begin{Soutput}
[1] 2.880557
\end{Soutput}
\begin{Sinput}
R> mean(abs(error)[results[,"sat"] == "SN2"], na.rm = TRUE)
\end{Sinput}
\begin{Soutput}
[1] 0.8527607
\end{Soutput}
\begin{Sinput}
R> cor(results$est, results$obs)
\end{Sinput}
\begin{Soutput}
[1] 0.9740032
\end{Soutput}
\end{Schunk}

The mean absolute error (MAE) of the estimates was 1.35 meters for both
satellites combined. Landsat-8 images led to higher errors (2.88 meters)
than Sentinel-2 (0.85 meters). The error from Sentinel-2 is closer to other
experiences (e.g., roughly 0.5 meters in \cite{ovakoglou:2016}) whereas
Landsat-8 errors are considerably larger. Potential sources of error are the
lower resolution of the satellite images affecting the detection of the
shoreline coupled with elevation errors triggered by the interpolation of the
topography. We consider that a thorough analysis of sources of error goes
beyond the scope of this manuscript.


\section{Summary and discussion} \label{sec:summary}

Satellite images are valuable and freely accessible sources of information 
provided by three major satellite programs: Landsat, MODIS and Copernicus.
Combining imagery from multiple programs can potentially improve the
spatio-temporal resolution of remotely sensed data. Formats, conventions, and
sharing protocols vary according to the satellite program, mission, and
data product, which may hinder data blending. 

Current \proglang{R} packages focus on single programs or specific tasks
concerning satellite images. We developed the \pkg{RGISTools} package as a mean
to access satellite data from multiple programs and from a single point. 
\pkg{RGistools} not only optimizes the access to the satellite images from
different programs using the more efficient APIs, but also it offers 
standardized functions for handling multi-program imagery. Additionally,
functions are designed to efficiently handle time-series from a computational
and memory standpoint. 
 
This manuscript begins with an overview of the package. The descriptions of the
workflow and the functionalities are coupled with a MODIS example that ends with
the application of the IMA statistical technique 
\citep{imaSmooth:2018, imaCloud:2019}for filling and smoothing satellite images.
Next, a case study shows intricacies of the package combining pre-processed
images from Landsat-8 and Sentinel-2 missions to estimate the water levels of
a reservoir in Northern Spain.

The package works locally with time-series of satellite images, which can be
challenging in memory terms (RAM and disk memory). The package uses three
strategies to address these challenges. It applies efficient routines such as
those in GDAL \citep{gdal:2019} whenever possible. It allows through functions
and arguments to remove unnecessary information for specific purposes. Images
are loaded in \proglang{R} at the end of the process, when images contain just
essential information for a specific task.

Moreover, we argue that working locally with satellite images is a sensible
option for statisticians and environmentalist that pursue the development of
new methods. \proglang{R} is a flexible environment to rapidly test tentative
methods and the eager evaluation enables the immediate assessment of the 
results. \proglang{R} is also an open source programming language which favors
a better understanding, application, and enhancement of existing spatio-temporal
methods. Working locally allows to benefit from these strengths at any point
of the workflow.

There is still room for improvement. \pkg{RGISTools} mainly deals with
satellite images as \class{Raster} class objects \citep{raster:2019}, which
is not straightforward when images are in various formats or heterogeneous.
Also, \class{Raster} objects only work with 3-dimensional arrays, which makes
it difficult to handle time-series of multispectral images since space, time
and spectral bands generally involve more than three dimensions. Packages under
development, such as \pkg{stars} \citep{stars:2019} and \pkg{gdalcubes} 
\citep{gdalcubes:2019} are promising solutions. \pkg{RGISGTools} already 
benefits from the computation advantages of \pkg{stars} to compute the remote
sensing indices, but its full integration depends on a thoroguh analysis that
is still pending. Finally, data fusion  techniques frequently involve radar
images \citep{dfreview:2019}. In its current version, the package downloads
radar images but does not give support to their processing. These and other
challenges that may arise in the future from more complex use cases, will be
resolved in subsequent versions of the package.


\section*{Computational details}

The results in this paper were obtained using
\proglang{R}~3.6.2 with the
\pkg{MASS}~7.3.51.4 package. \proglang{R} itself
and all packages used are available from the Comprehensive
\proglang{R} Archive Network (CRAN) at
\url{https://CRAN.R-project.org/}.

\section*{Acknowledgments}
This research was supported by the project MTM2017-82553-R (AEI/FEDER,
UE). It has also received funding from la Caixa Foundation (ID1000010434),
Caja Navarra Foundation and UNED Pamplona, under agreement LCF/PR/PR15/51100007.


\bibliography{refs}

%
%
%
%
%


\end{document}